\documentclass[twocolumn,aps,showpacs,nofootinbib,prd]{revtex4-1}
\usepackage{amssymb}
\usepackage{amsmath}
\usepackage{graphics}
\usepackage{epsfig}
\usepackage[dvips]{color}
\usepackage{float}
\newcommand{\be}{\begin{equation}}
\newcommand{\ee}{\end{equation}}
\newcommand{\bea}{\begin{eqnarray}}
\newcommand{\eea}{\end{eqnarray}}
\newcommand{\beas}{\begin{eqnarray*}}
\newcommand{\eeas}{\end{eqnarray*}}

\newcommand{\slsh}[1]{{\not \! #1}}

\newcommand{\dpi}{(2\pi)}
\newcommand{\p}{\parallel}
\newcommand{\pp}{\perp}
\newcommand{\ga}{\gamma_{\alpha}}
\newcommand{\gm}{\gamma_{\mu}}
\newcommand{\gn}{\gamma_{\nu}}

\newcommand{\gmn}{g^{\mu\nu}}
\newcommand{\gma}{g^{\mu\alpha}}
\newcommand{\gan}{g^{\alpha\nu}}

\begin{document}
\title{Prompt photon yield and elliptic flow from gluon fusion induced by magnetic fields in relativistic heavy-ion collisions}
\author{Alejandro Ayala$^{1,2}$, Jorge David Casta\~no-Yepes$^1$, C. A. Dominguez$^2$,\\ L. A. Hern\'andez$^1$, Sa\'ul Hern\'andez-Ortiz$^1$, Mar\1a Elena Tejeda-Yeomans$^3$.}
  \address{
  $^1$Instituto de Ciencias
  Nucleares, Universidad Nacional Aut\'onoma de M\'exico, Apartado
  Postal 70-543, M\'exico Distrito Federal 04510,
  Mexico.\\
  $^2$Centre for Theoretical and Mathematical Physics, and Department of Physics,
  University of Cape Town, Rondebosch 7700, South Africa.\\
  $^3$Departamento de F\'{\i}sica, Universidad de Sonora, Boulevard Luis Encinas J. y Rosales, Colonia
  Centro, Hermosillo, Sonora 83000, Mexico.}

\begin{abstract}
We compute photon production at early times in semi-central relativistic heavy-ion collisions from non-equilibrium gluon fusion induced by a magnetic field. The calculation accounts for the main features of the collision at these early times, namely, the intense magnetic field and the high gluon occupation number. The gluon fusion channel is made possible by the magnetic field and would otherwise be forbidden due to charge conjugation invariance. Thus, the photon yield from this process is an excess over calculations without magnetic field effects. We compare this excess to the difference between PHENIX data and recent hydrodynamic calculations for the photon transverse momentum distribution and elliptic flow coefficient $v_2$. We show that with reasonable values for the saturation scale and magnetic field strength, the calculation helps to better describe the experimental results obtained at RHIC energies for the lowest part of the transverse photon momentum. 

\end{abstract}

\maketitle

Heavy-ion reactions carried out at the BNL Relativistic Heavy-Ion Collider (RHIC) and at the CERN Large Hadron Collider (LHC) produce magnetic fields with an initial intensity in the interaction region estimated to be as high as several times the mass of the pion squared~\cite{intensity}. Due to event-by-event fluctuations of protons in the target and the projectile, it has been shown~\cite{Bzdak} that the magnitude of the magnetic field can be of the same order in central and in peripheral collisions. Nevertheless, the field strength reaches its highest values for non-central collisions. Although event-by-event, the intensities of both of the components transverse to the beam axis are comparable, the mean of the component along the reaction plane is centered at vanishing intensity.  Therefore, unless event-by-event observables are studied, only the field component perpendicular to the reaction plane needs to be accounted for. These intense fields are however short lived and they fade out fast with time. 

A magnetic field makes possible the production of photons from processes otherwise not allowed. For instance, it has been shown that the QCD trace anomaly can turn the energy momentum of the soft gluon bulk into photons~\cite{Skokov}. Photons can also be emitted by  magnetic field induced bremsstrahlung and pair annihilation in the quark-gluon plasma (QGP)~\cite{Zakharov}. In addition, quarks can emit photons by synchrotron radiation~\cite{Tuchin}. Other approaches to study photon production in the presence of an intense magnetic field include the gauge/gravity correspondence in a strongly coupled ${\mathcal{N}}= 4$ plasma~\cite{Leonardo}. These novel calculations have recently been implemented  to try explain the experimentally measured excess~\cite{experimentsyield, experimentsv2,experiments2} of photons over models that describe well other low momentum observables. The enhanced production of photons and their azimuthal anisotropy in heavy-ion reactions has also been studied in the absence of magnetic field effects, {\it e.g.} from the deceleration processes of the two colliding nuclei~\cite{Kodama}, from the modification of the quark and gluon distributions~\cite{McLerran}, from glasma induced processes~\cite{Larry2, Raju3} and from the delayed formation of the QGP~\cite{Liu}.

A magnetic field naturally produces an asymmetry in the emission of electromagnetic radiation since it provides a direction that breaks translational symmetry. Therefore, magnetic fields can be a source of not only an excess in the photon yield, but also of the puzzling large strength of the coefficient $v_2$ in the Fourier expansion of the azimuthal distribution. The latter has been found to be as large as that of pions~\cite{v2photons}. Although some recently improved hydrodynamic \cite{hydro-photons1,hydro-photons2} and transport~\cite{transport} calculations obtain a better agreement with ALICE and PHENIX measurements of low and intermediate transverse momentum photons, this agreement is not yet complete~\cite{review}. Therefore, it remains important to quantify the fraction of the yield, and of the asymmetry arising from magnetic field effects, if any, to better characterize the initial stages of heavy-ion reactions.

Notice in addition that it is also at the earliest times of a high-energy heavy-ion collision that the largest gluon occupation number is achieved, due to the shattering of the over-ocupied initial state called the {\it glasma}~\cite{Larry2}.  It is then natural to explore a mechanism where collisions of these non-equilibrated gluons induce the emission of photons accounting at the same time for the presence of an intense magnetic field. 

In this work we compute the production of prompt photons from the perturbative fusion of low momentum gluons coming from the shattered glasma, which are highly abundant early in the collision due to saturation effects, that are important  from times of order $\tau_s\sim 1/\Lambda_s$, where $\Lambda_s$ is the saturation scale~\cite{Lappi}, up to a time $\Delta\tau_s \simeq 1.5$ fm~\cite{Larry2}. Since the intensity of the magnetic field decreases fast with time, here we assume that the field's magnitude can be represented by a single value, though during a short early period of time of order $\Delta \tau_s$. This is a simplification that leaves room for improvement \cite{Deng}. A similar approach, albeit in the context of thermalized gluons and with a series of simplifying approximations was attempted in Ref.~\cite{Ayala}. Here we perform a more complete calculation without resorting to assuming early gluon thermalization and without kinematical restrictions.

Since the presence of a magnetic field breaks translational invariance, the amplitude for the process has to be computed in coordinate space and subsequently integrated over space-time. The lowest order process in the strong, $\alpha_s=g^2/4\pi$, and electromagnetic, $\alpha_{\tiny{em}}=e^2/4\pi$, couplings come from an amplitude made out of a quark {\it triangle} diagram with two gluons and one photon attached each at one of the vertices of the triangle. As stated in Refs.~\cite{Larry2} and~\cite{Raju3}, the over occupied gluon state can be described as made out of quasi-particles, therefore perturbative methods are applicable.  

The quark propagator is written in its coordinate space representation as~\cite{Schwinger}
\bea
   S(x,x')=\Phi (x,x')\int\frac{d^4p}{(2\pi)^4}e^{-ip\cdot (x-x')}S(p)\;,
\label{genprop}
\eea
where
\bea
   \!\!\!\!\!\!\!\!\!\Phi (x,x')=\exp\left\{i|q_f|\int_{x'}^xd\xi^\mu\left[A_\mu + \frac{1}{2}
   F_{\mu\nu}(\xi - x')^\nu\right]\right\}\!\!,
\label{phase}
\eea
is called the {\it phase factor} and $q_f$ is the quark's charge. We consider the contribution of three light flavors, thus $q_u = 2e/3$ and $q_d = q_s = e/3$. 

The Fourier transform of the translationally invariant part of the propagator is given by
\begin{eqnarray}
   && i S(p) = \int _{0}^{\infty}\frac{d\tau}{\cos(|q_fB|\tau)}
   e^{i\tau(p_{\parallel}^{2} - p_{\perp}^{2}\frac{\tan (|q_fB|\tau)}{|q_fB|\tau}- m_f^{2} +i\epsilon)}
   \times\nonumber\\
   && \!\!\!\!\biggl[\!\left(\cos (|q_fB|\tau)\!\! + \!\! \gamma_1 \gamma_2 \sin (|q_fB|\tau)\right)
   \!\!(m_f+\slsh{p_{\|}})\!\!  - \!\!  \frac{\slsh{p_\bot}}{\cos(|q_fB|\tau)} \biggr]\! ,\nonumber\\
   \label{tracewithSchwinger}
 \end{eqnarray}
where $m_f$ is the quark mass. We have chosen the homogeneous magnetic field to point in the $\hat{z}$ direction, namely $\boldsymbol{B}=B\hat{z}$. This configuration can be obtained from an external vector potential which we choose in the so called {\it symmetric gauge} $A^{\mu}= \frac{B}{2}(0,-y,x,0)$. We have also defined $
p_\perp^\mu\equiv(0,p_1,p_2,0),\
p_\parallel^\mu\equiv(p_0,0,0,p_3),\
p_\perp^2\equiv p_1^2+p_2^2$ and
$p_\parallel^2\equiv p_0^2-p_3^2$, and therefore $p^2=p_\parallel^2 - p_\perp^2$.

The expression for the amplitude is given by 
\bea
   \widetilde{{\mathcal{M}}}&=&-\int \!d^4xd^4yd^4z\int\!\frac{d^4r}{(2\pi)^4}
   \frac{d^4s}{(2\pi)^4}\frac{d^4t}{(2\pi)^4}
   \nonumber\\
   &\times&e^{-it\cdot (y-x)}e^{-is\cdot (x-z)}e^{-ir\cdot (z-y)}e^{-ip\cdot z}e^{-ik\cdot y}e^{iq\cdot x}\nonumber\\
   &\times&
   \Big\{
   {\mbox{Tr}}\left[ iq_f\gamma_\alpha iS(s) ig\gamma_\mu t^c iS(r) ig\gamma_\nu t^d iS(t) \right]
   \nonumber\\
   &+&
   {\mbox{Tr}}\left[ iq_f\gamma_\alpha iS(t) ig\gamma_\nu t^d iS(r) ig\gamma_\mu t^c iS(s) \right]
   \Big\}
   \nonumber\\
   &\times&\Phi(x,y)\Phi(y,z)\Phi(z,x)\epsilon^\mu(\lambda_p)\epsilon^\nu(\lambda_k)\epsilon^{\alpha}(\lambda_q),
   \label{amplitude}
\eea
where the space-time points $z$, $y$ and $x$ correspond to the vertices where the gluons with four-momenta $p$ and $k$ (Lorentz indices $\mu$ and $\nu$) and the photon with four-momentum $q$ (Lorentz index $\alpha$) are attached, respectively. The factors $t^c$, $t^d$ are Gell-Mann matrices. The polarizations vectors for the gluons and the photon are $\epsilon^{\mu}(\lambda_p)$, $\epsilon^{\nu}(\lambda_k)$, and $\epsilon^{\alpha}(\lambda_q)$, respectively. The two traces correspond to the two possible ways the charge flows in the triangle. 

The product of phase factors can be written as
\bea
   \Phi(x,y)\Phi(y,z)\Phi(z,x)
   =e^{i\frac{|q_fB|}{2}\epsilon_{ij}(z-x)_i(x-y)_j},
   \label{prodphase}
\eea
where the indices $i,j=1,2$ correspond to the transverse components of the corresponding vectors and $\epsilon_{ij}$ is the Levi-Civita symbol. We used the explicit form of $A^\mu$ which gives $F_{12}=-F_{21}=-B$, with the rest of the components of $F_{\mu\nu}$ vanishing. The integrations over space-time points in Eq.~(\ref{amplitude}) are carried out more easily by making the change of variables $\omega=z-x$ and $l=x-y$, after which one gets
\bea
   &&\int \!d^4xd^4yd^4z\Phi(x,y)\Phi(y,z)\Phi(z,x)\nonumber\\
   &\times&e^{-ir\cdot (y-x)}e^{-is\cdot (z-y)}e^{-it\cdot (x-z)}e^{-ip\cdot z}e^{-ik\cdot y}e^{iq\cdot x}\nonumber\\
   &=&(2\pi)^4\delta^4(q-k-p)
   \nonumber\\
   &\times&
   \int d^4\omega d^4le^{-i\omega\cdot (r-s+p)}e^{-il\cdot (r-t-k)}e^{i\frac{|q_fB|}{2}\epsilon_{ij}\omega_i l_j},
   \label{afterchange}
\eea
which exhibits the overall energy-momentum conservation in the process.

We now use the fact that when the magnetic field is very intense, as compared to the other energy (squared) scales involved in the computation of this amplitude, the quark dynamics is dominated by the lowest Landau levels. For the case of non-thermal quarks, this means that the magnetic field is taken to satisfy $eB\gg m_f^2$. We thus hereby set $m_f=0$.

\begin{figure}[t]
\begin{center}
\includegraphics[scale=.5]{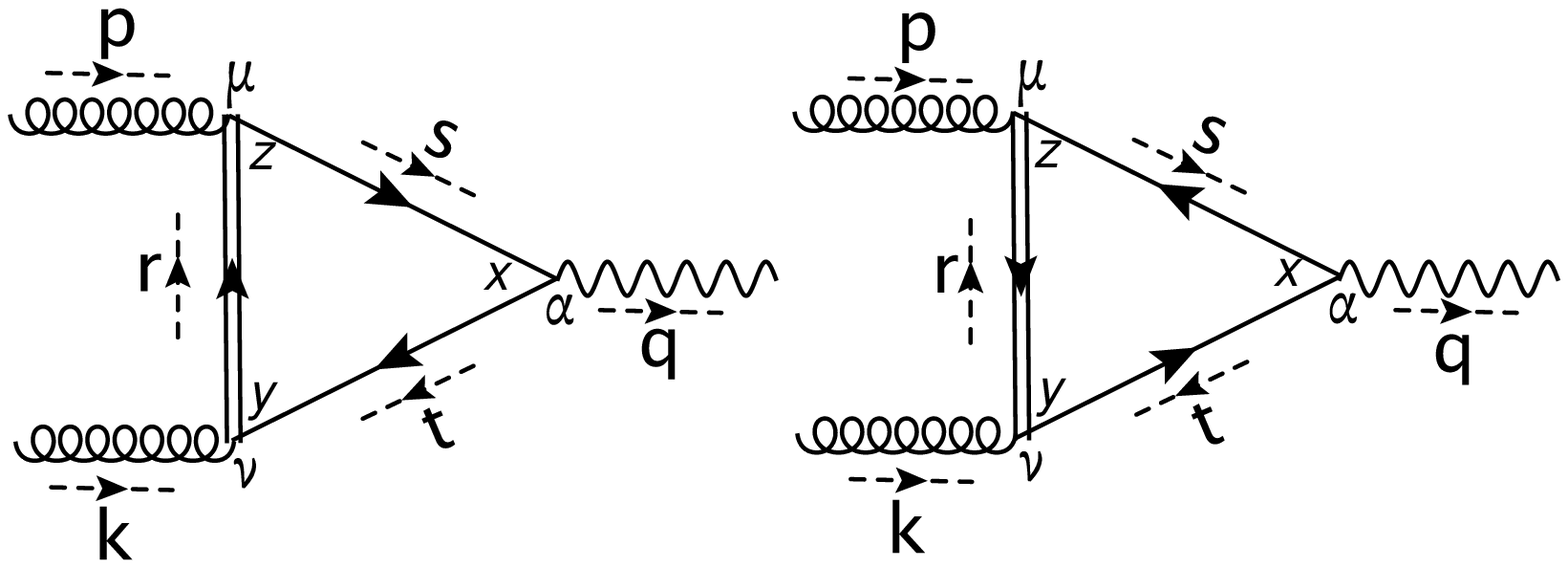}\\
\includegraphics[scale=.5]{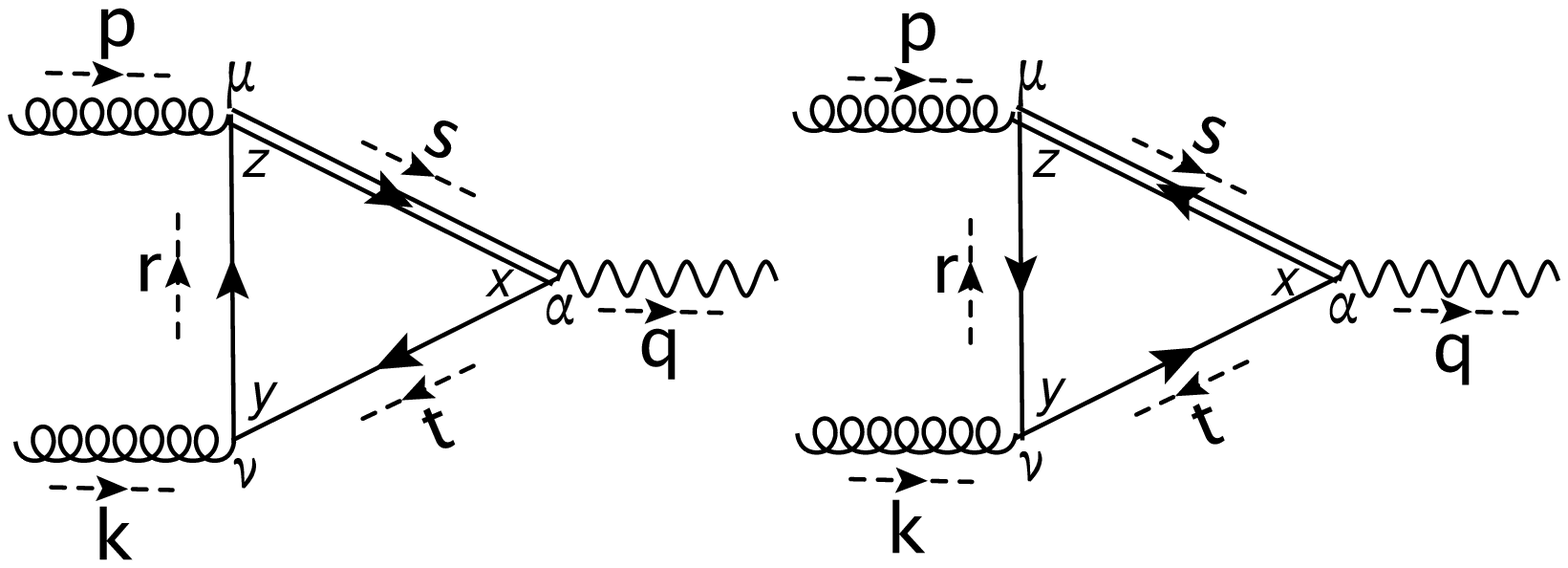}\\
\includegraphics[scale=.5]{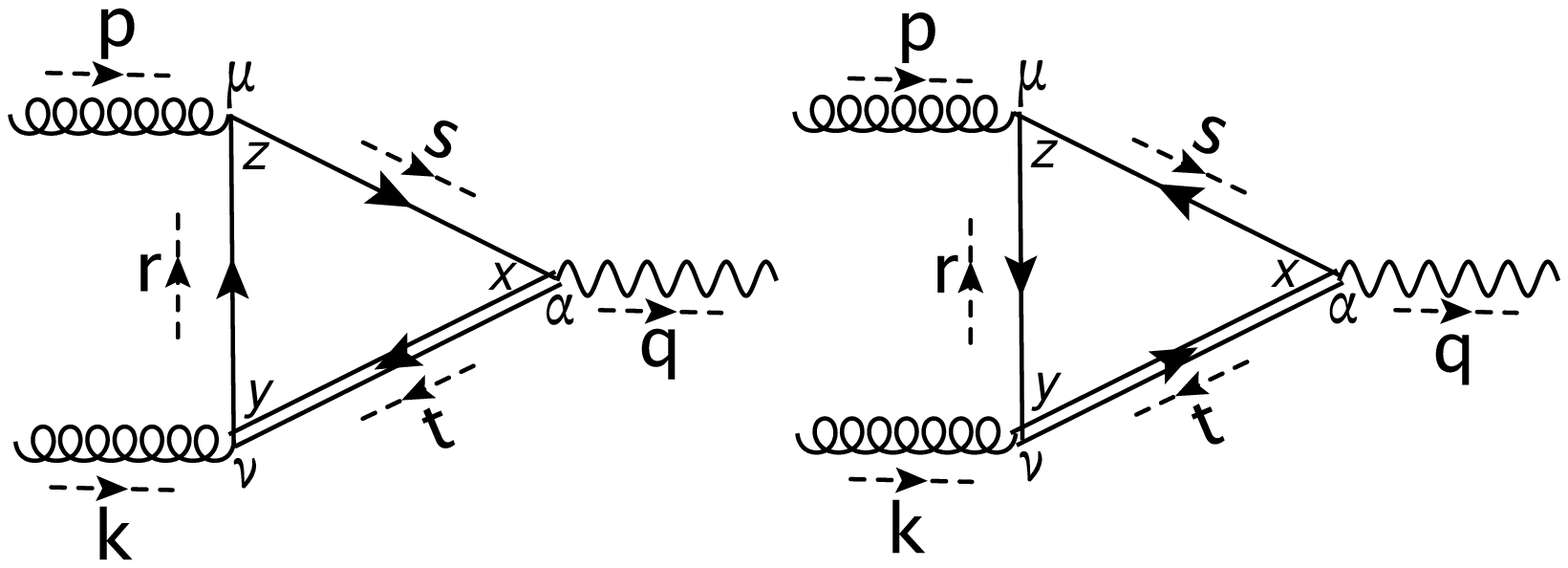}
\caption{Feynman diagrams representing the dominant contribution for magnetic field gluon fusion induced photon emission. The double lines represent that the corresponding propagator is in the 1LL. The arrows in the propagators represent the direction of the flow of charge. The arrows at the sides of the propagator lines represent the momentum direction.}
\label{fig1}
\vspace{-1cm}
\end{center}
\end{figure}

For the lowest (LLL) and the first excited (1LL) Landau levels, the corresponding propagators can explicitly be written as~\cite{Miransky}
\bea
   S^{\mbox{\small{LLL}}}(p)&=&-2ie^{-\frac{p_\perp^2}{|q_fB|}}\frac{\slsh{p_{\parallel}}}{p_\parallel^2}
   {\mathcal{O}}^+_\parallel \nonumber\\
   S^{\mbox{\small{1LL}}}(p)&=&\frac{e^{-\frac{p_\perp^2}{|q_fB|}}}{p_\parallel^2-2|q_fB|}\nonumber\\
   &\times&\left\{ \slsh{p_{\parallel}} {\mathcal{O}}^+_\parallel
   \left[ 1- \frac{2p_\perp^2}{|q_fB|}  \right] - \slsh{p_{\parallel}} 
   {\mathcal{O}}^-_\parallel + 4 \slsh{p_{\perp}} \right\}.
   \label{propLLL}
\eea
The operators ${\mathcal{O}}^\pm_\parallel=\left[1\pm ({\mbox{sign}}(q_fB)) i\gamma_1\gamma_2 \right]/2$ project onto the longitudinal space. It can be shown that when the three propagator lines each contain an operator ${\mathcal{O}}^\pm_\parallel$, the amplitude $\widetilde{{\mathcal{M}}}$ vanishes. This means that, in order to consider the dominant contribution for magnetic field gluon fusion induced photon emission, one of the quark propagators needs to be in the 1LL and the other two in the LLL. Selection rules of this sort have been discussed in the context of photon splitting in magnetic fields in Ref.~\cite{Adler}. The amplitude for the process is depicted in Fig.~\ref{fig1}, where the double lines are meant to represent that the corresponding propagator is in the 1LL. The amplitude becomes
\bea
\widetilde{{\mathcal{M}}}&=&8i\dpi^4\delta^{(4)}\left(q-k-p\right)
\delta^{cd}|q_f| g^2\nonumber\\
&\times&\int\frac{d^4r}{\dpi^4}\frac{d^4s}{\dpi^4}\frac{d^4t }{\dpi^4}\nonumber\\
&\times&\int d^4w\;d^4l e^{-il(r-t-k)}e^{-iw(r-s+p)}\nonumber\\
&\times&\exp\left\{-i\frac{|q_fB|}{2}\epsilon_{mj}w_ml_j\right\}\exp\left\{-\frac{r_\pp^2+s_\pp^2+t_\pp^2}{|q_fB|}\right\}\nonumber\\
&\times&\text{Tr}\left\{\frac{\gamma_1\gamma_2\ga\slsh{t}_\pp\gn\slsh{r}_\p\gm^{\p}\slsh{s}_\p}{r_\p^2s_\p^2\left(t_\p^2-2\left| q_fB\right|\right)} +\frac{\gamma_1\gamma_2\gm\slsh{s}_\pp\ga\slsh{t}_\p\gn^{\p}\slsh{r}_\p}{t_\p^2r_\p^2\left(s_\p^2-2\left| q_fB\right|\right)}\right.\nonumber\\
&+&\left.\frac{\gamma_1\gamma_2\gn\slsh{r}_\pp\gm\slsh{s}_\p\ga^{\p}\slsh{t}_\p}{s_\p^2t_\p^2\left(r_\p^2-2\left| q_fB\right|\right)}\right\}\epsilon^{\mu}(\lambda_p)\epsilon^\nu(\lambda_k)\epsilon^{\alpha}(\lambda_q).
\label{matrixelem}
\eea
To carry out the integrations in Eq.~(\ref{matrixelem}), we simplify the denominators considering that $2|q_fB|\gg t_\p^2,\ s_\p^2,\ r_\p^2$. Recall that the square of the parallel components of a vector $p^\mu$ is given by  $p^2_\parallel=p_0^2-p_3^2$, this difference is small when looking at central rapidity, thus the approximation. Since the majority of the gluons in the shattered glasma have momenta much less than $\Lambda_s$, this is a good working approximation. After a lengthy but straightforward calculation, the matrix element can be written as
\bea
\!\!\!\!\widetilde{{\mathcal{M}}}&=&-i\dpi^4\delta^{(4)}\left(q-k-p\right)
\frac{|q_f| g^2\delta^{cd}e^{f\left(p_\pp,k_\pp\right)}}{32\pi\dpi^8}\nonumber\\
&\times&\left\{\left(\frac{1}{2}\gma_\p-\frac{p^{\mu}_\p p^{\alpha}_\p}{p^2_\p}\right)h^{\nu}(a)-\left(\frac{1}{2}\gmn_\p-\frac{p^{\mu}_\p p^{\nu}_\p}{p^2_\p}\right)h^{\alpha}(a)\right.\nonumber\\
&+&\left(\frac{1}{2}\gmn_\p-\frac{k^{\mu}_\p k^{\nu}_\p}{k^2_\p}\right)h^{\alpha}(b)-\left(\frac{1}{2}\gan_\p-\frac{k^{\alpha}_\p k^{\nu}_\p}{k^2_\p}\right)h^{\mu}(b)\nonumber\\
&+&\left.\left(\frac{1}{2}\gan_\p-\frac{q^{\alpha}_\p q^{\nu}_\p}{q^2_\p}\right)h^{\mu}(c)-\left(\frac{1}{2}\gma_\p-\frac{q^{\mu}_\p q^{\alpha}_\p}{q^2_\p}\right)h^{\nu}(c)\right\}\nonumber\\
&\times&\epsilon^{\mu}(\lambda_p)\epsilon^\nu(\lambda_k)\epsilon^{\alpha}(\lambda_q),
\label{matrixelemafterapprox}
\eea
with $h^{\mu}(a)=(i/\pi)\epsilon_{ij}a^ig^{j\mu}_\pp$, $a_i=p_i + 2k_i - i\epsilon_{im}p_m$, $b_i=2p_i - k_i - i\epsilon_{im}k_m$, $c_i=p_i - k_i + i\epsilon_{im}(p_m + k_m)$ and
\bea
f\left(p_\pp,k_\pp\right)&=&\frac{1}{8|q_fB|}\left(p_m-k_m+i\epsilon_{mj}(p_j+k_j)\right)^2\nonumber\\
&-&\frac{1}{2|q_fB|}\left(p_m^2+k_m^2+2i\epsilon_{jm}p_mk_j\right),
\eea
where $g_\pp={\mbox{diag}}(1,1)$ and $g_\parallel={\mbox{diag}}(1,-1)$ are the metric tensors in the transverse and longitudinal spaces.
\begin{figure}[t!]
\begin{center}
\includegraphics[scale=.45]{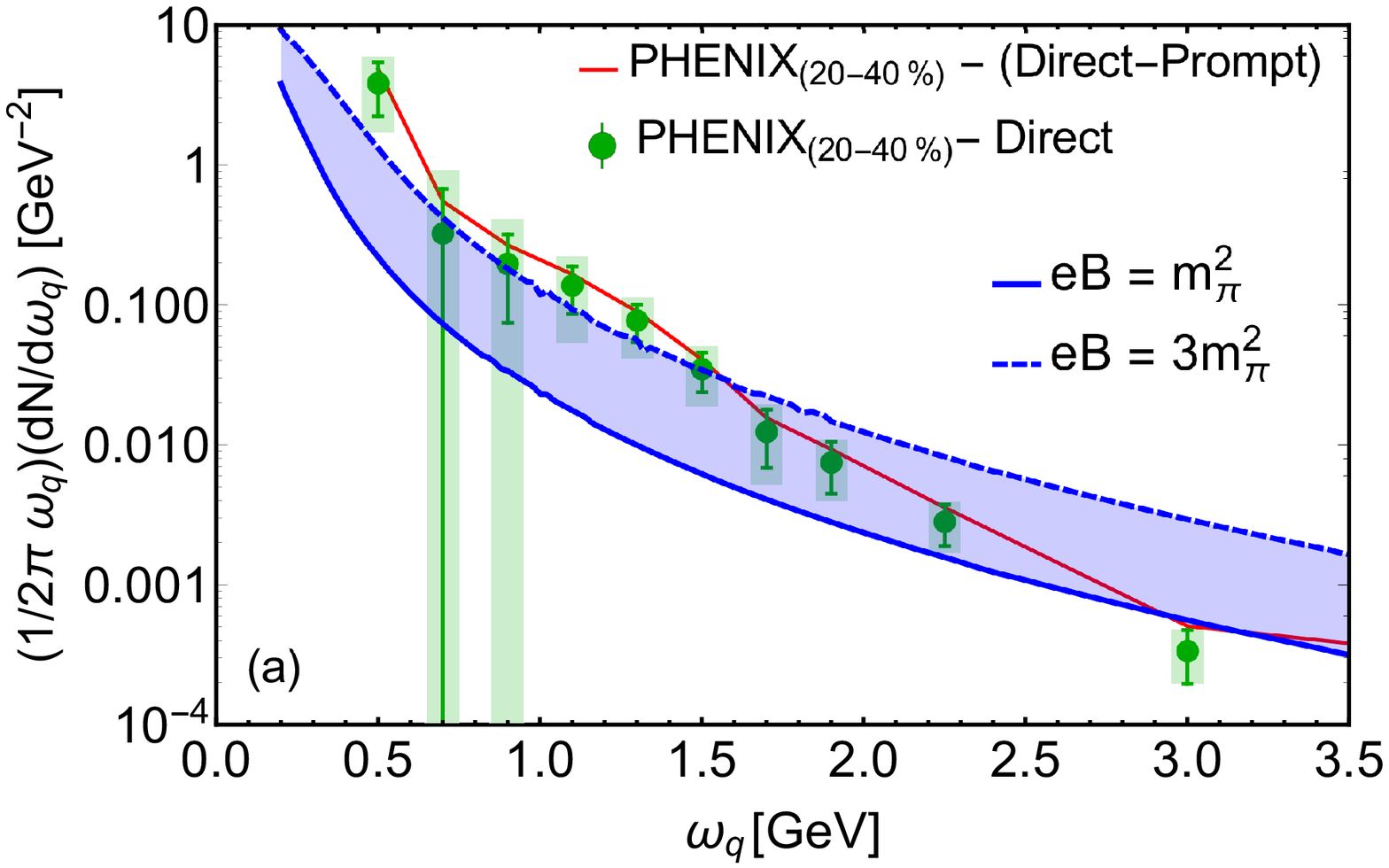}
\end{center}
\begin{center}
\includegraphics[scale=.45]{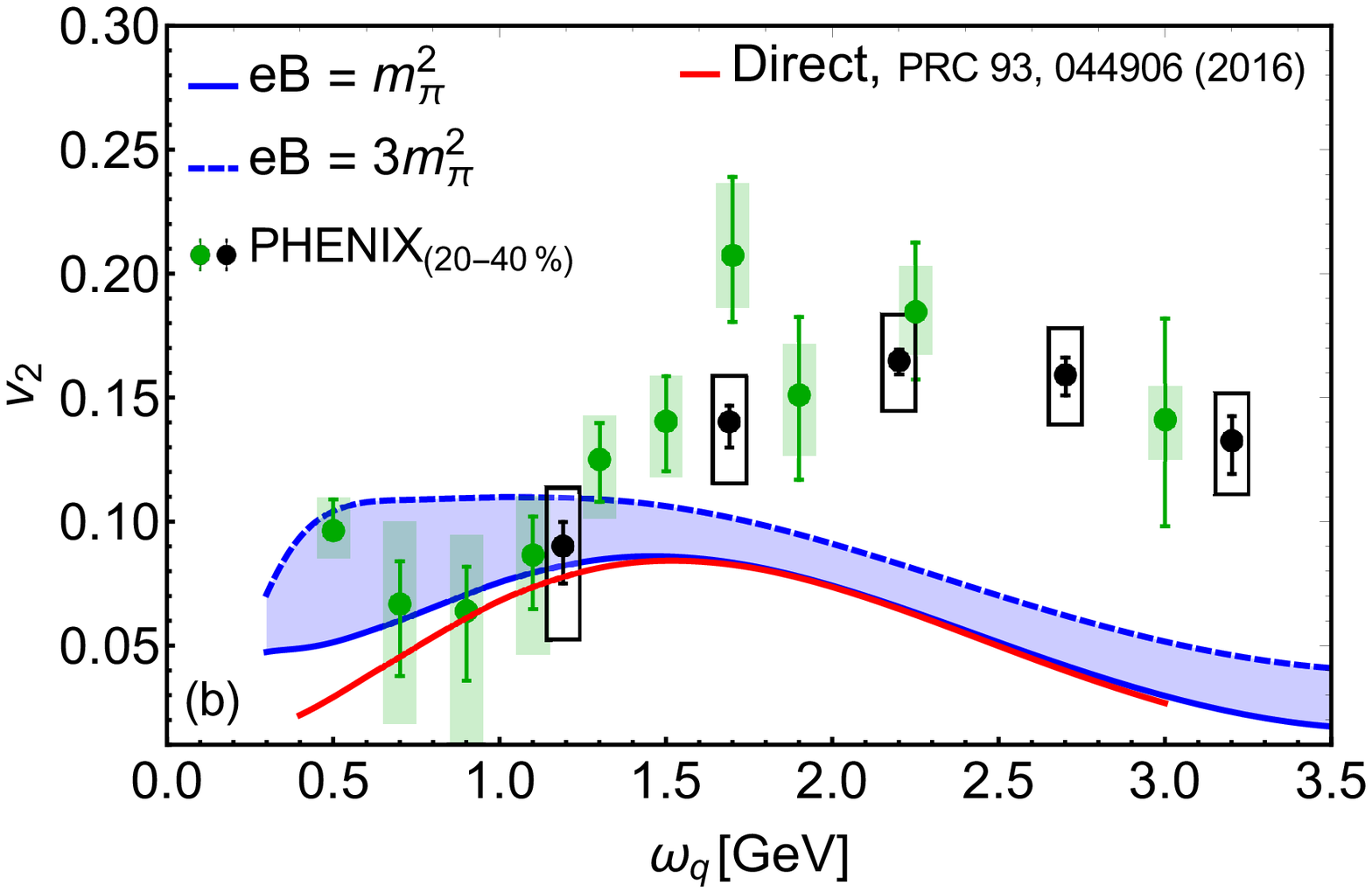}
\end{center}
\caption{(a) Difference between PHENIX photon invariant momentum distribution~\cite{experimentsyield} and direct (points) or direct minus prompt (zigzag) photons from Ref.~\cite{hydro-photons1} compared to the yield from the present calculation. (b) Harmonic coefficient $v_2$ combining the calculation of Ref.~\cite{hydro-photons1} and the present calculation compared to PHENIX data~\cite{experimentsv2}. Curves are shown as functions of the photon energy for central rapidity and the centrality range 20-40\%. Only the experimental error bars are shown. The bands show variations of the parameter $eB$ within the indicated ranges and computed with $\alpha_s=0.3$, $\Lambda_s=2$ GeV, $\eta=3$, $\Delta \tau_s=1.5$ fm, $R=7$ fm, $\beta=0.25$ and $\chi=0.8$.}
\label{fig2}
\end{figure}
Equation~(\ref{matrixelemafterapprox}) represents the leading order matrix element contribution in $q_fB$ to photon production from gluon fusion in the shattered glasma. 
Notice that this matrix element is not transverse, in the sense that when one replaces only one or two of the three polarization vectors by the corresponding momenta, the contraction does not yield a vanishing result. 
Nevertheless, it can be checked that if the three polarization vectors are replaced by their corresponding momenta, one does obtain a vanishing result. This is a nontrivial property of the above matrix element which arises because one of the three propagators is in the 1LL.

In order to find the photon production probability we square the matrix element in Eq.~(\ref{matrixelemafterapprox}) and average over initial gluon and sum over final photon polarizations
\bea
   \!\!\!\!\!\frac{1}{4}\sum_{\mbox{\small{pol}}}|\widetilde{{\mathcal{M}}}|^2= \dpi^4\delta^{(4)}\left(q-k-p\right) 
   {\mathcal{V}}\tau_s\frac{1}{4}\sum_{\mbox{\small{pol}}}|{\mathcal{M}}|^2,
\eea
where ${\mathcal{V}}\tau_s$ is the space-time volume where the process takes place. We find explicitly
\bea
   \frac{1}{4}\sum_{\mbox{\small{pol}}}
   |{\mathcal{M}}|^2&=&\frac{q_f^2\alpha_{\mbox{\tiny{em}}}\alpha_s^2}{(2\pi)
   \omega_q^2}\left(\omega_p^2+3\omega_k^2\right)q_\pp^2
   \nonumber\\
   &\times&
   \exp\left\{-\frac{q_\pp^2}{q_fB\omega_q^2}\left[\omega_p^2+\omega_k^2-\omega_p\omega_k\right]\right\}\!.
   \label{matrixwotilde}
\eea
In writing Eq.~(\ref{matrixwotilde}) we have already used that, in order to satisfy energy and momentum conservation for massless gluons and photons, the four-momenta $p^\mu=(\omega_p,\vec{p})$, $k^\mu=(\omega_k,\vec{k})$ and $q^\mu=(\omega_q,\vec{q})$ satisfy
\bea
   p^\mu&=&\omega_p(1,\hat{p})
   =\left(\omega_p/\omega_q\right)q^\mu,\nonumber\\
   k^\mu&=&\omega_k(1,\hat{k})
   =\left(\omega_k/\omega_q\right)q^\mu,
\eea
which means that for the reaction to take place, both gluons are required to have parallel momenta and the produced photon to move in the original gluons' direction. When the mediumÕs dispersive properties are accounted for and thus the magnetic field-dependent refraction index is included, gluons and photons are not collinear anymore.  A calculation of the effect of such dispersive properties is currently being performed and will soon be reported elsewhere.

The invariant photon momentum distribution is thus given by
\bea
\omega_q\frac{dN^{\mbox{\tiny{mag}}}}{d^3q}&=&\frac{\chi {\mathcal{V}} \Delta \tau_s}{2(2\pi)^3}
\int\frac{d^3p}{\dpi^32\omega_p}\int\frac{d^3k}{\dpi^32\omega_k}
n(\omega_p)n(\omega_k)\nonumber\\
&\times&\dpi^4\delta^{(4)}\left(q-k-p\right)\frac{1}{4}\sum_{\mbox{\small{pol}},f}|{\mathcal{M}}|^2,
\label{invdist}
\eea
where we have included the sum over the three considered flavors and $n(\omega)$ represents the distribution of gluons coming from the shattered glasma. Also, we have introduced a factor $\chi$, to account for the fact that the overlap region in a semi-central collision is not the full nuclear volume. We use for this distribution a simple model that accounts for the high occupation gluon number given by~\cite{Larry2, Krasnitz}
\bea
   n(\omega)=\frac{\eta}{e^{\omega/\Lambda_s}-1},
\eea
where $\eta$ represents the high gluon occupation factor and $\Lambda_s$ is, as before, the saturation scale. Also, in order to account for the sudden change in pressure between the interaction region and vacuum when the glasma is shattered, we introduced a {\it flow velocity} factor whose effect is to shift the gluon energies in the product of the matrix element with the gluon distributions, that is, $\omega_{p,k}\to (p,k)\cdot u$. For simplicity we allow for a constant {\it flow} velocity $u^\mu=\gamma(1,\beta)$, with $\gamma=1/\sqrt{1-\beta^2}$. A flow velocity that can be developed during the early stages of the collision even in out-of equilibrium processes has been discussed previously, for instance, in Ref.~\cite{Larry2}.   

\begin{figure}[t!]
\begin{center}
\includegraphics[scale=.45]{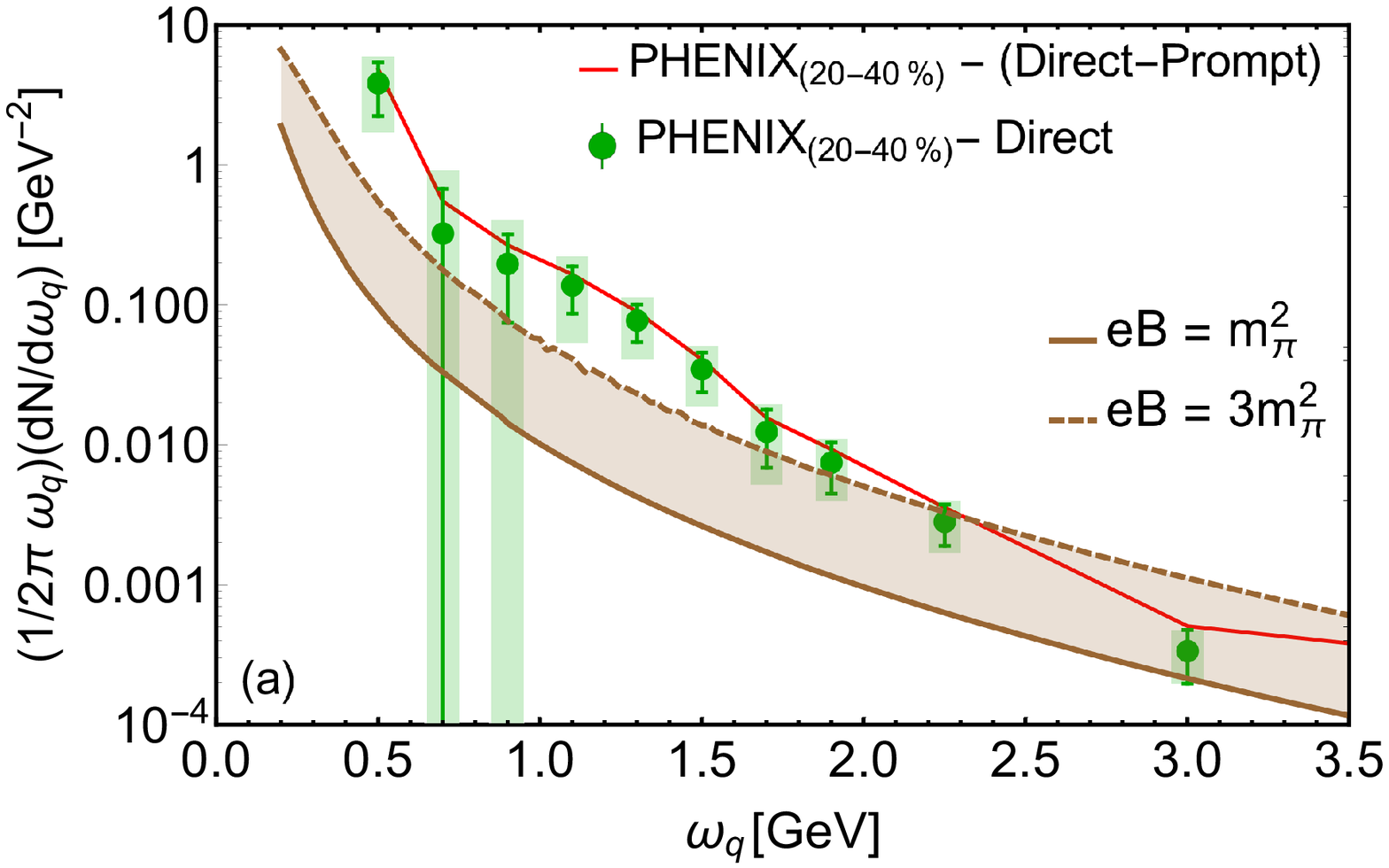}
\end{center}
\begin{center}
\includegraphics[scale=.45]{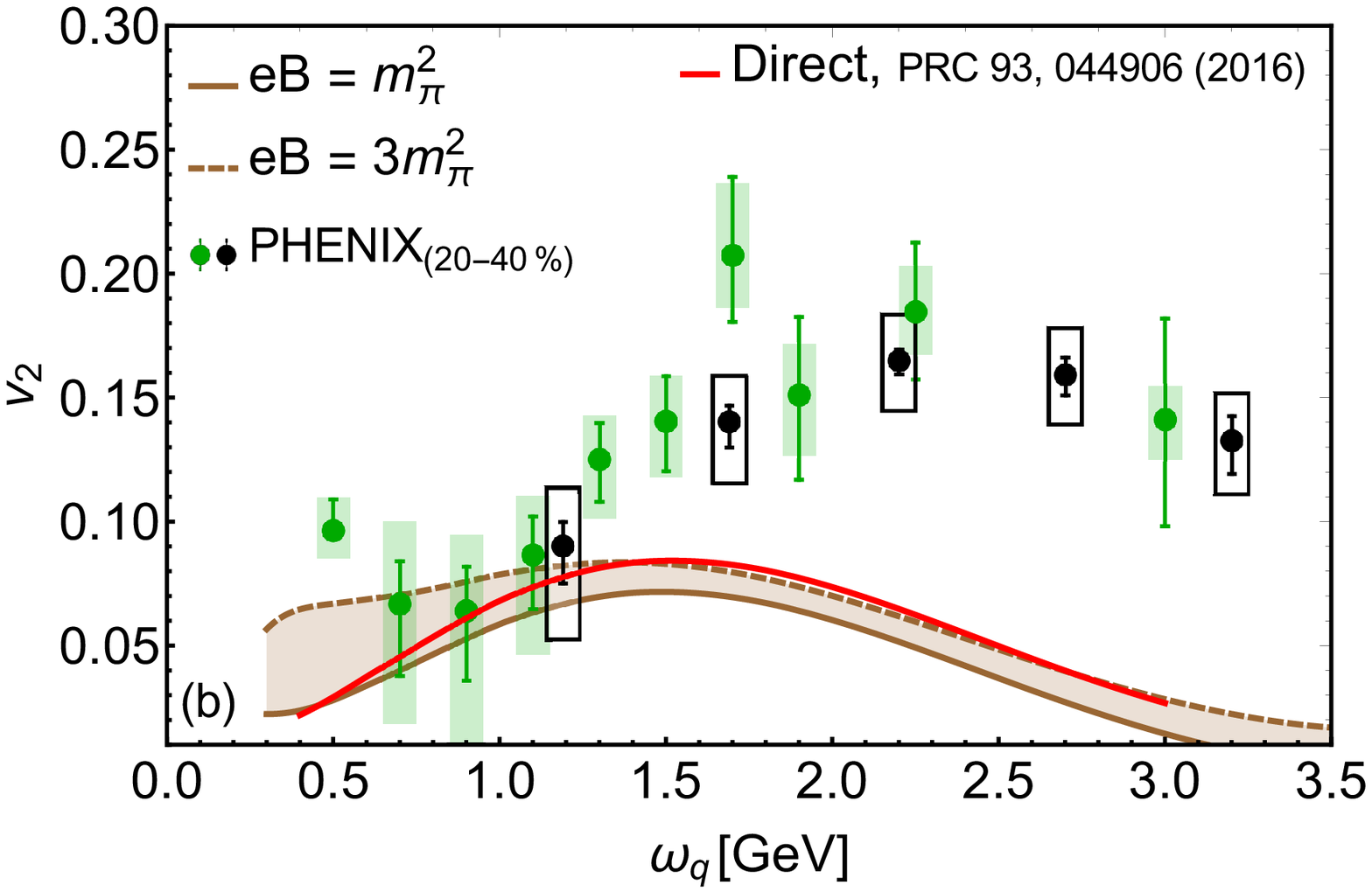}
\end{center}
\caption{(a) Difference between PHENIX photon invariant momentum distribution~\cite{experimentsyield} and direct (points) or direct minus prompt (zigzag) photons from Ref.~\cite{hydro-photons1} compared to the yield from the present calculation. (b) Harmonic coefficient $v_2$ combining the calculation of Ref.~\cite{hydro-photons1} and the present calculation compared to PHENIX data~\cite{experimentsv2}. Curves are computed with the same set of parameters as Fig.~\ref{fig2} but with $\beta=0$.}
\label{fig3}
\end{figure}

In order to explicitly compute the photon distribution and the coefficient $v_2$, recall that the magnitude of the photon's momentum transverse (to de direction of the magnetic field), $q_\perp$, is obtained projecting the magnitude of the photon momentum with $\sin (\theta)$, where $\theta$ is the angle between the magnetic field direction and the photon direction of motion. In order to refer $q_\perp$ to the reaction plane, we use that $\sin(\theta) = \sin(\pi/2 - \phi)=\cos(\phi)$, where $\phi$ is the angle between the photon's momentum and the reaction plane. The azimuthal distribution with respect to the reaction plane can be given in terms of a Fourier decomposition as
\bea
   \frac{dN^{\mbox{\tiny{mag}}}}{d\phi}=\frac{N^{\mbox{\tiny{mag}}}}{2\pi}
   \left[1+\sum_{i=1}^\infty 2v_n(\omega_q)\cos(n\phi)\right],
\label{Fourier}
\eea
from where, together with Eq.~(\ref{invdist}), the coefficient $v_2(\omega_q)$ can be extracted. The total number of photons, $N^{\mbox{\tiny{mag}}}$ is
\bea
   N^{\mbox{\tiny{mag}}}=\int \frac{d^3q}{(2\pi)^3} \frac{dN^{\mbox{\tiny{mag}}}}{d^3q}.
\eea
Figure~\ref{fig2}(a) shows the difference between PHENIX data~\cite{experimentsyield} for the invariant momentum distribution and the state-of-the-art hydrodynamical calculation of Ref.~\cite{hydro-photons1}. The points represent the difference between PHENIX data and direct photons. To get a rough estimate of the uncertainty in this subtraction, we also show with the zigzag curve the difference between PHENIX and direct minus prompt photons. Figure~\ref{fig2}(b) shows the harmonic coefficient $v_2$, using the direct photon result of Ref.~\cite{hydro-photons1} together with our calculation, also compared to PHENIX data~\cite{experimentsv2}. The curves are shown as functions of the photon energy for central rapidity and the centrality range 20-40\%. Only the experimental error bars are shown. Notice that $v_2$ is a weighed average namely,
\bea
   v_2(\omega_q)=
   \frac{
   \frac{dN^{\mbox{\tiny{mag}}}}{d\omega_q}(\omega_q)\
   v_2^{\mbox{\tiny{mag}}}(\omega_q)
   +
   \frac{dN^{\mbox{\tiny{direct}}}}{d\omega_q}(\omega_q)\
   v_2^{\mbox{\tiny{direct}}}(\omega_q)} 
   {\frac{dN^{\mbox{\tiny{mag}}}}{d\omega_q}(\omega_q) 
   + 
   \frac{dN^{\mbox{\tiny{direct}}}}{d\omega_q}(\omega_q)},\nonumber\\
\eea
where $dN^{\mbox{\tiny{direct}}}/d\omega_q$ and $v_2^{\mbox{\tiny{direct}}}$ are the ($\omega_q$-dependent) spectrum and second harmonic coefficient of direct photons from Ref.~\cite{hydro-photons1}, respectively. For the calculations we work with $\alpha_s=0.3$, $(g_s=2)$, $\Lambda_s=2$ GeV, $\eta=3$, $\Delta \tau_s=1.5$ fm, ${\mathcal{V}}=(4\pi R^3)/3$, with $R=7$ fm (corresponding to the Au nuclear radius), $\beta=0.25$, $\chi=0.8$ and a field intensity $eB$ in the range $m_\pi^2<eB<3m_\pi^2$, which corresponds to the values at  $\tau \simeq 0.5$ fm. To quantify the systematic effects of the flow factor, Fig.~\ref{fig3} shows the calculation with the same set of parameters as in Fig.~\ref{fig2} but with $\beta=0$. Notice that the calculation without the flow factor becomes shallower for higher energies than when this factor is included, both for the yield and $v_2$. This is to be expected since flow produces a blue shift in the spectra. The absence of the flow factor also produces a less steep rise of $v_2$ for small energies. 

As can be seen from Figs.~\ref{fig2} and~\ref{fig3}, the excess photon yield and $v_2$ coming from magnetic field induced gluon fusion helps to better describe the experimental data having as a baseline a state-of-the-art calculation accounting for many of the well described sources of photons. The effect on the photon yield is to increase the distribution and at the same time shift it to higher photon energy values. For the case of $v_2$, the agreement of the calculation with data is particularly good in the lowest part of the spectrum since it describes well the observed experimental fall between 0.5 and 1 GeV. This fall has received little attention and in our approach it is due to the rise and fall of the $v_2$ distribution that peaks for energy values of the order of $\sqrt{eB}$. For the energy region above 1 GeV the calculation falls short of data. This may be due to the fact that the gluon distribution we used does not contain a power-like tail which is known to better describe the numerical solutions for this kind of distribution~\cite{Krasnitz}.

The picture that emerges is as follows: In a semi-central high-energy heavy-ion collision, a magnetic field of a large intensity is produced. The time scales when this field is the most intense are also the scales associated to the production of a large number of small momentum gluons coming from the shattering of the glasma. The magnetic field provides the mechanism to allow that these shattered gluons fuse and convert into photons in excess over other well studied mechanisms that may happen during the entire evolution of the system. The magnetic field also provides an initial asymmetry for the development of an azimuthal anisotropy quantified in terms of a substantial $v_2$, particularly at low photon momenta. The spectrum and the azimuthal anisotropy are a bit hardened if one considers a simple expansion scenario whose physical origin is the sudden change in pressure for the liberated glue from within the interaction region and the outside vacuum. 

In conclusion, we have shown that perturbative gluon fusion is a channel opened by the presence of a magnetic field (of a realistic intensity) at the early stages of a high-energy heavy ion collision, that can contribute to better describe the yield and $v_2$ of photons at the lowest end of the spectrum.  A more detailed systematic study of the calculation here shown is in preparation and will be soon reported elsewhere.

The authors are in debt with C. Gale and J.-F. Paquet for providing us with the numbers of their calculation and for useful comments and suggestions. The authors also thank M. Greif for useful comments. Support for this work has been received in part by UNAM-DGAPA-PAPIIT grant number IN101515, by Consejo Nacional de Ciencia y Tecnolog\1a grant number 256494 and by NRF (SouthAfrica) and the Research Administration University of Cape Town.

\end{document}